\documentclass{PoS}
\usepackage{amsmath}
\usepackage{xspace}
\usepackage{nicefrac}

\newcommand{\Tevatron}{{\sc Tevatron}\xspace}
\newcommand{\CDF}{CDF\xspace}
\newcommand{\Dzero}{D\O\ }

\newcommand{\MCatNLO}{M\protect\scalebox{0.8}{C}@N\protect\scalebox{0.8}{LO}\xspace}

\newcommand{\NLOPS}{N\protect\scalebox{0.8}{LO}P\protect\scalebox{0.8}{S}\xspace}

\newcommand{\MEPSatLO}{M\protect\scalebox{0.8}{E}P\protect\scalebox{0.8}{S}@L\protect\scalebox{0.8}{O}\xspace}
\newcommand{\MEPSatNLO}{M\protect\scalebox{0.8}{E}P\protect\scalebox{0.8}{S}@N\protect\scalebox{0.8}{LO}\xspace}
\newcommand{\LO}{LO\xspace}
\newcommand{\NLO}{NLO\xspace}

\newcommand{\Sherpa}{S\protect\scalebox{0.8}{HERPA}\xspace}
\newcommand{\Comix}{C\protect\scalebox{0.8}{OMIX}\xspace}

\newcommand{\Amegic}{A\protect\scalebox{0.8}{MEGIC++}\xspace}
\newcommand{\CSS}{C\protect\scalebox{0.8}{SS}\xspace}

\newcommand{\GoSam}{G\protect\scalebox{0.8}{O}S\protect\scalebox{0.8}{AM}\xspace}

\newcommand{\done}{\mathrm{d}}

\newcommand{\va}{\vphantom{\int_A^B}}
\newcommand{\vb}{\vphantom{\dfrac{\va}{\va}}}

\title{\vspace{-3.5cm}{\small IPPP/13/93, DCPT/13/186, MPP--2013--293, 
                                   MCNET--13--19, LPN13--092, SLAC--PUB--15812}\\\vspace{2cm}
       NLO merging in $\mathbf{t\bar{t}}$+jets}

\ShortTitle{NLO merging in $t\bar{t}$+jets}

\author{\speaker{Marek Sch\"onherr}\\
        Institute for Particle Physics Phenomenology,
        Durham University, Durham DH1 3LE, UK}

\author{Stefan H\"oche\\
        SLAC National Accelerator Laboratory, 
        Menlo Park, CA 94025, USA}

\author{Junwu Huang\\
        SLAC National Accelerator Laboratory, 
        Menlo Park, CA 94025, USA}

\author{Gionata Luisoni\\
        Max-Planck Institut f{\"u}r Physik,
        F{\"o}hringer Ring 6, 80805 M{\"u}nchen, Germany}

\author{Jan Winter\\
        Max-Planck Institut f{\"u}r Physik,
        F{\"o}hringer Ring 6, 80805 M{\"u}nchen, Germany}

\abstract{In this talk the application of the recently introduced methods 
          to merge NLO calculations of successive jet multiplicities to the 
          production of top pairs in association with jets will be discussed, 
          in particular a fresh look is taken at the top quark forward-backward 
          asymmetries. 
          Emphasis will be put on the achieved theoretical accuracy and the 
          associated perturbative and non-perturbative error estimates.} 

\FullConference{The European Physical Society Conference on High Energy Physics -EPS-HEP2013\\
		18-24 July 2013\\
		Stockholm, Sweden}

\begin{document}

\section{Introduction}
\label{sec:intro}

The forward--backward asymmetry of an observable $O$ in top-quark pair 
production, as measured by the \CDF and \Dzero experiments at the 
$p\bar{p}$ collider \Tevatron \cite{Abazov:2007ab,Aaltonen:2008hc,
  Aaltonen:2011kc,Abazov:2011rq,Aaltonen:2012it} is defined as
\begin{equation}
  A_\text{FB}(O)
  \;=\;\dfrac{\hspace*{2mm}\left.\dfrac{\done\sigma_{t\bar{t}}}{\done O}\right|_{\Delta y>0}
	      -\left.\dfrac{\done\sigma_{t\bar{t}}}{\done O}\right|_{\Delta y<0}\hspace*{2mm}}
	     {\vb\left.\dfrac{\done\sigma_{t\bar{t}}}{\done O}\right|_{\Delta y>0}
	      +\left.\dfrac{\done\sigma_{t\bar{t}}}{\done O}\right|_{\Delta y<0}}
\end{equation}
where $\Delta y=y_t-y_{\bar t}$ is the rapidity difference 
between the top and the antitop quark. In both inclusive and differential 
asymmetry measurements unexpectedly large deviations from the Standard 
Model predictions were found.
Besides triggering substantial investigations of beyond-the-Standard-Model 
theories, higher order corrections in the Standard Model were calculated. Of 
particular importance is the influence of the parton shower, investigated in 
\cite{Skands:2012mm}, as it captures effects indispensable for experimental 
measurements. Further, all Monte Carlo event generators which are currently 
being used by experiments provide at most the inclusive production of 
$t\bar{t}$-pairs at \NLOPS accuracy. While \NLOPS matched calculations 
of $t\bar{t}$+jet production have been available \cite{Alioli:2011as,Kardos:2011qa} 
for a long time, they have not been combined with the inclusive simulation of 
$t\bar{t}$ production allowing improved predictions of $A_{\rm FB}$. 
\cite{Hoeche:2013mua} remedied this situation, providing a merged simulation 
of $t\bar{t}$ and $t\bar{t}+$jet production at hadron colliders, which 
preserves both the NLO accuracy of the fixed-order prediction and the 
logarithmic accuracy of the parton shower. Thus, accurate predictions for 
both the transverse momentum dependent asymmetry above a certain threshold 
and the inclusive asymmetries can be made. Electroweak corrections, 
calculated in \cite{Hollik:2011ps}, are not included.

\section{Results}
\label{sec:results}

In this study the \Sherpa \cite{Gleisberg:2008ta} event generator with 
its internal matrix element generators \Amegic \cite{Krauss:2001iv,
  Gleisberg:2003ue} and \Comix \cite{Gleisberg:2008fv}, its 
Catani-Seymour/Catani-Dittmaier-Seymour-Trocsanyi \cite{Catani:1996vz,
  Catani:2002hc} dipole subtraction \cite{Gleisberg:2007md}, and its 
\CSS parton shower \cite{Schumann:2007mg} has been used. The one-loop 
matrix element provided by the publicly available \GoSam package 
\cite{Cullen:2011ac,Mastrolia:2010nb} have been interfaced through the 
Binoth-Les-Houches accord (BLHA) \cite{Binoth:2010xt,Alioli:2013nda}. 
The MSTW2008 LO/NLO PDF sets \cite{Martin:2009iq} are employed for the 
leading order merged (\MEPSatLO \cite{Hoeche:2009rj,Hoeche:2010kg}) and 
next-to-leading order merged (\MEPSatNLO \cite{Hoeche:2012yf,Gehrmann:2012yg}) 
calculations, respectively. 

For the \MEPSatNLO calculation the $p\bar{p}\to t\bar{t}$ and 
$p\bar{p}\to t\bar{t}j$ processes, calculated at next-to-leading 
order accuracy and matched to the parton shower individually 
using the variant of the \MCatNLO technique described in 
\cite{Hoeche:2011fd},\footnote{The applicability of this method to 
processes of most general colour structures was demonstrated in 
\cite{Hoeche:2012ft,Hoeche:2012fm,Cascioli:2013era}.} have been merged 
as prescribed in \cite{Hoeche:2012yf,Gehrmann:2012yg}. 
The scales in these calculations are set according to the prescription 
given therein, i.e.\
\begin{equation}
  \begin{split}
   \alpha_s^{n+k}(\mu_R)
   \,=\;\alpha_s^n(\mu_\text{core})\;\prod\limits_{i=1}^k\alpha_s(t_i)\;,
  \end{split}
\end{equation}
where the $t_i$ are the emission scales identified in the backwards 
clustering, and $\mu_\text{core}$ is a freely chosen scale for the 
identified core process \cite{Hoeche:2009rj}. 
Two central scale choices have been investigated: 
$\mu_\text{core}=m_{t\bar{t}}$ and $\mu_\text{core}=\mu_\text{QCD}=
2\,\left|p_ip_j\right|$ ($i,j$ large-$N_c$ colour partners). Thus, 
$\mu_\text{QCD}$ is a scale inspired by the colour flow of the event.

\begin{figure}[t!]
  \begin{center}
    \includegraphics[width=0.33\textwidth]{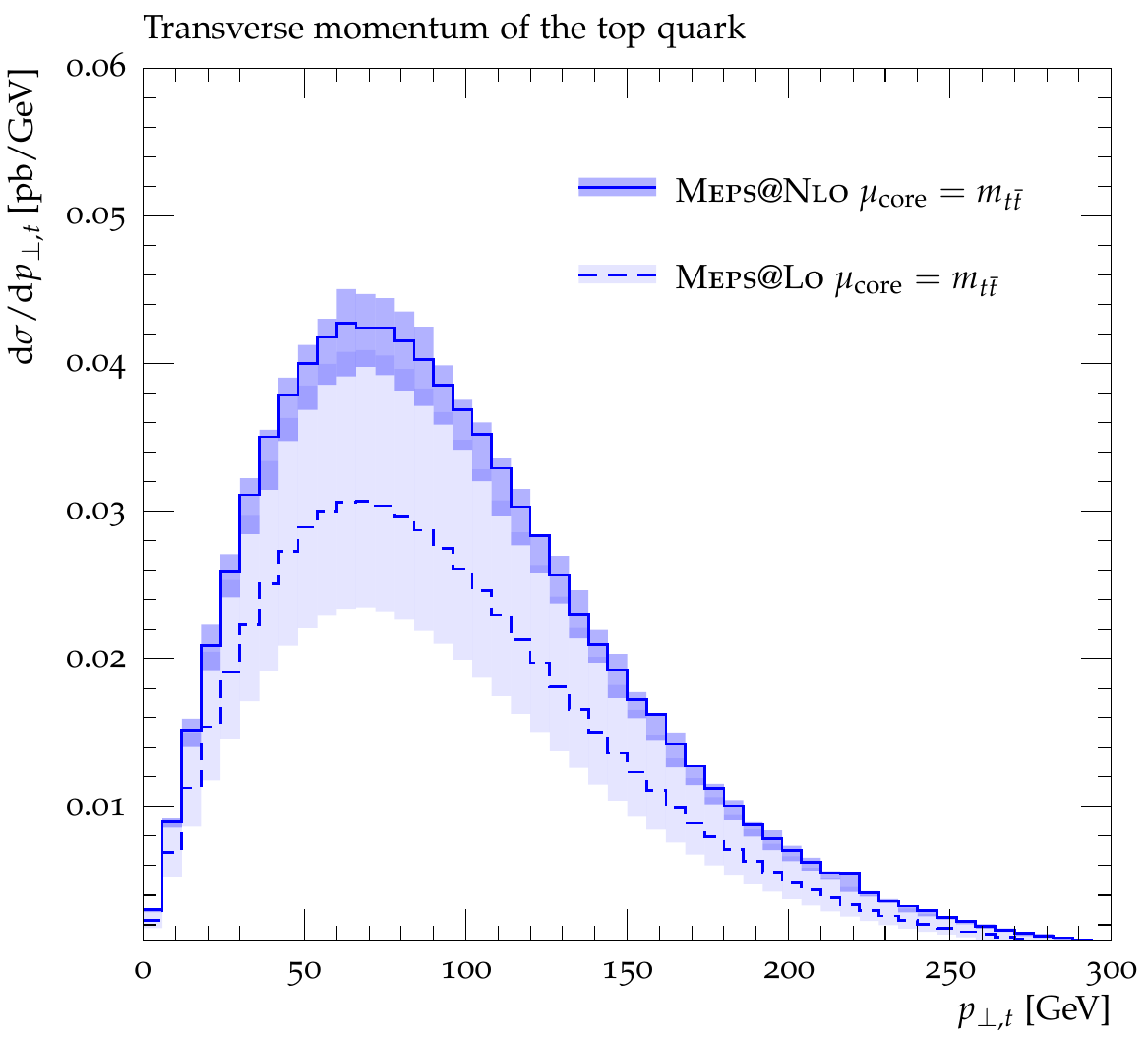}\hfill
    \includegraphics[width=0.33\textwidth]{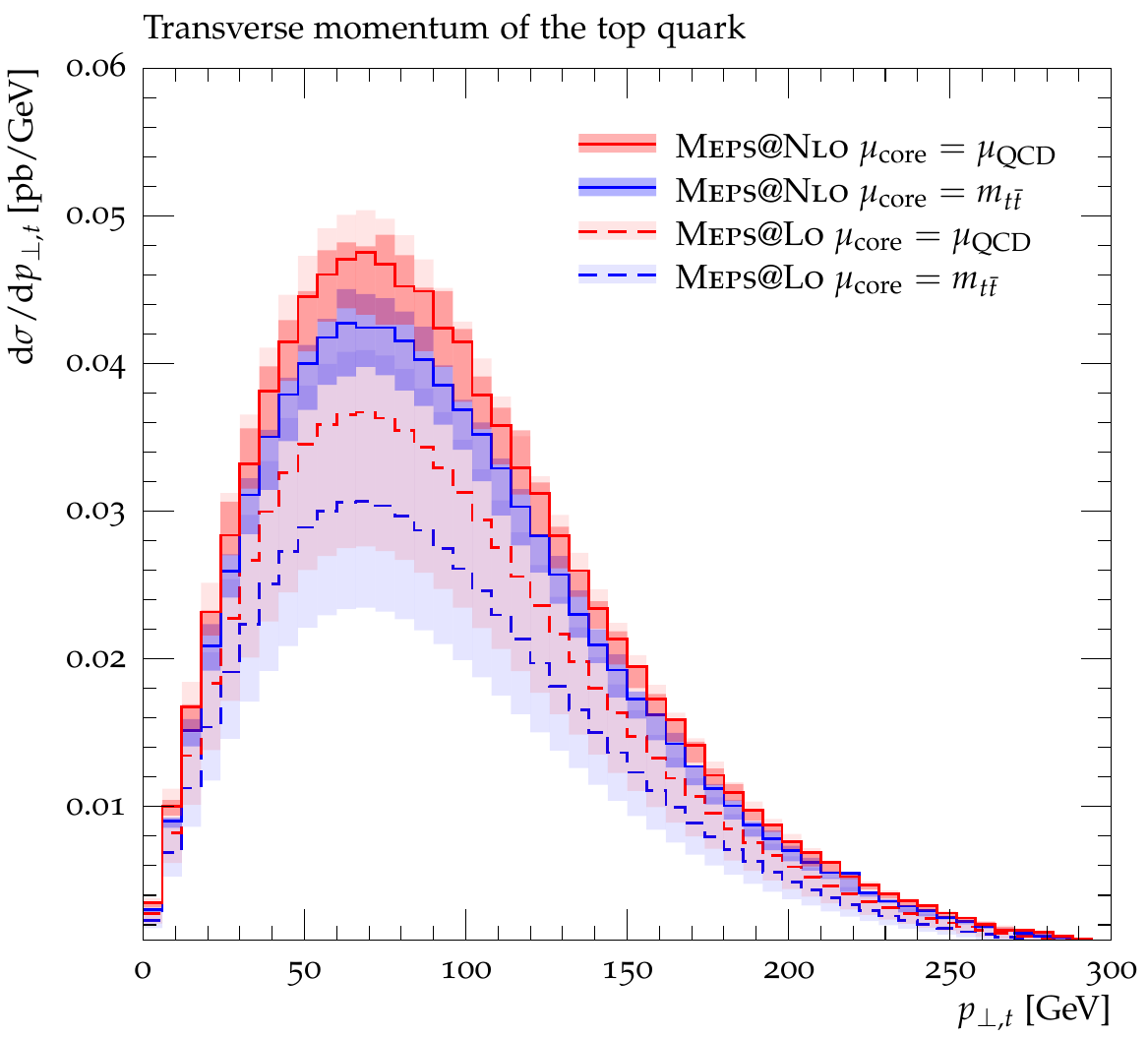}\hfill
    \includegraphics[width=0.33\textwidth]{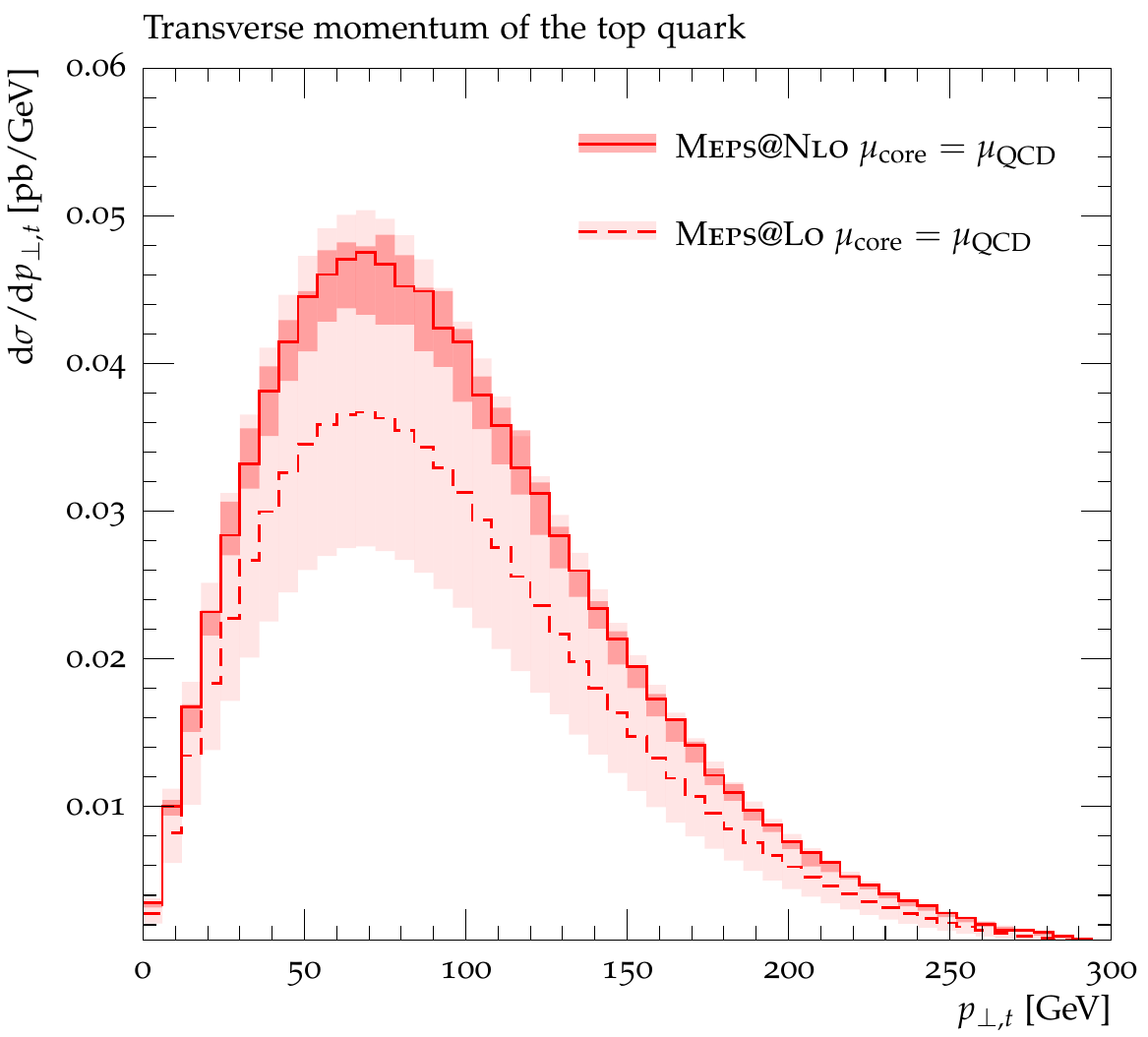}
  \end{center}
  \vspace*{-5mm}
  \caption
  {
    The scale dependence of the transverse momentum of the top quark in 
    top quark pair production at the \Tevatron for two different choices 
    of core scales, $\mu_\text{core}=m_{t\bar{t}}$ (left) and 
    $\mu_\text{core}=\mu_\text{QCD}$ (right). The customary 
    scale uncertainties are indicated by the dark (\NLO) and light (\LO) 
    coloured bands, respectively.
    The central figure combines the two outer figures for easy comparison. 
    \label{fig:afb-scales}
  }
  \vspace*{-2mm}
\end{figure}

\begin{figure}[t]
  \begin{center}
    \includegraphics[width=0.47\textwidth]{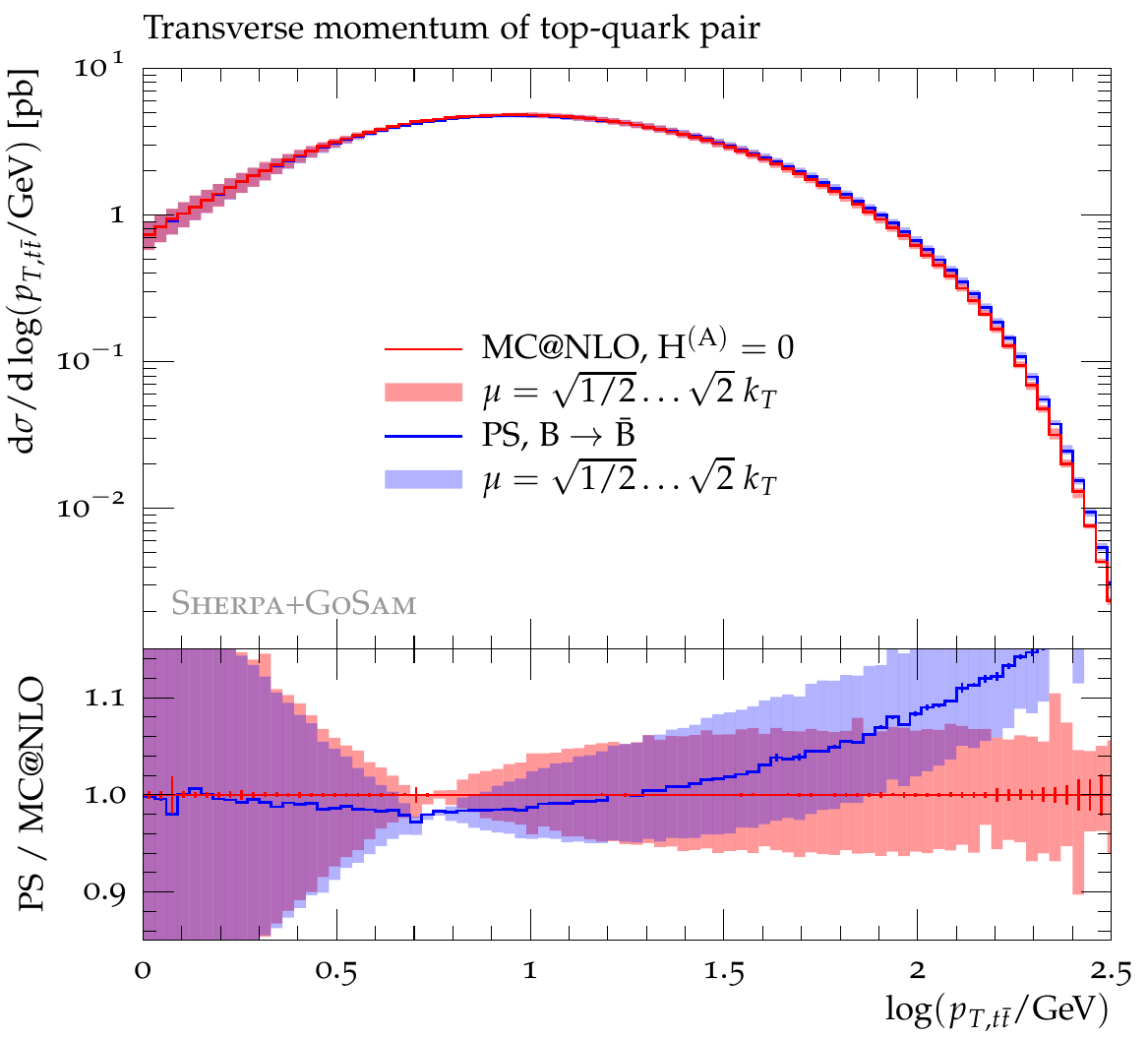}\hfill
    \includegraphics[width=0.47\textwidth]{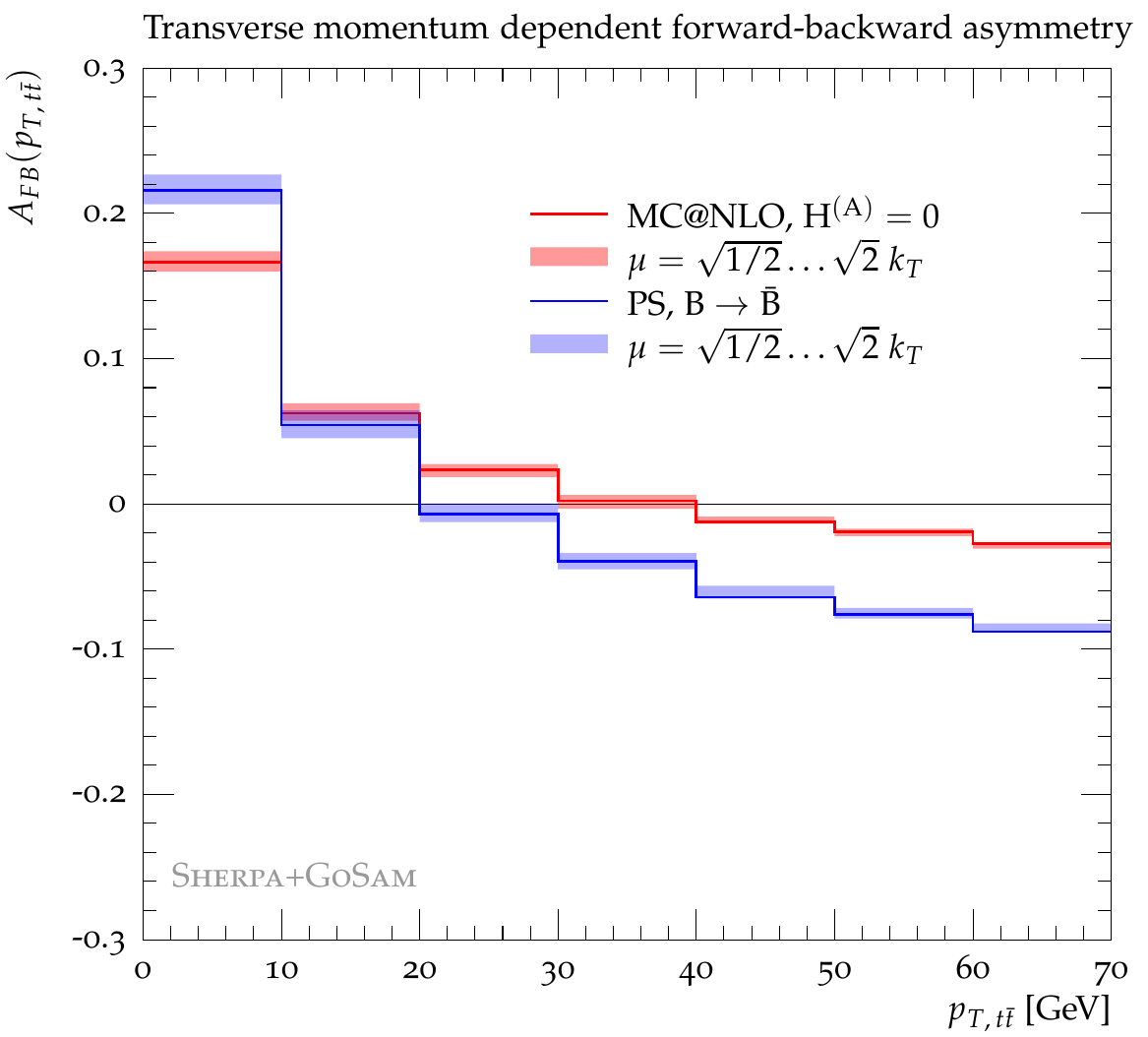}
  \end{center}
  \vspace*{-5mm}
  \caption
  {
    The importance of $N_c=3$ soft-gluon colour-coherence on the 
    transverse momentum of the $t\bar{t}$-pair (left) and the associated 
    forward-backward asymmetry (right). The uncertainties are shown, 
    stemming from a variation of the renormalisation scale in the 
    resummation kernels of either the parton shower kernels (blue) or the 
    subleading-colour improved \MCatNLO kernels (blue). The parton shower 
    is supplied with a local $K$-factor to compensate for the different 
    normalisations. Likewise, the \MCatNLO calculation is deprived of the 
    fixed-order real emission correction, thus the resulting predictions 
    differ only in subleading colour terms.
    \label{fig:afb-coherence}
  }
  \vspace*{-3mm}
\end{figure}

\begin{figure}[t]
  \begin{center}
    \includegraphics[width=0.47\textwidth]{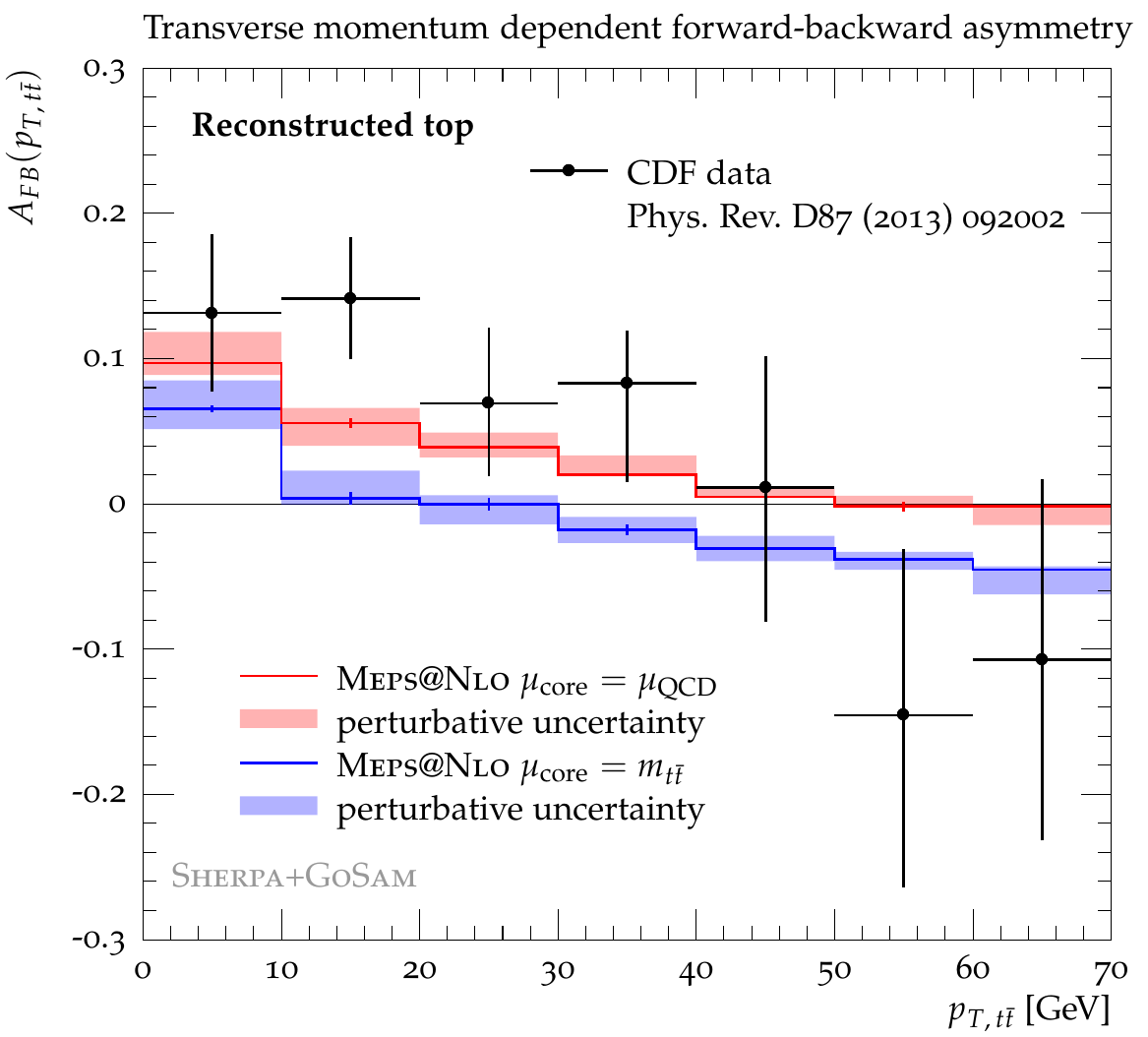}\\\vspace*{2mm}
    \includegraphics[width=0.47\textwidth]{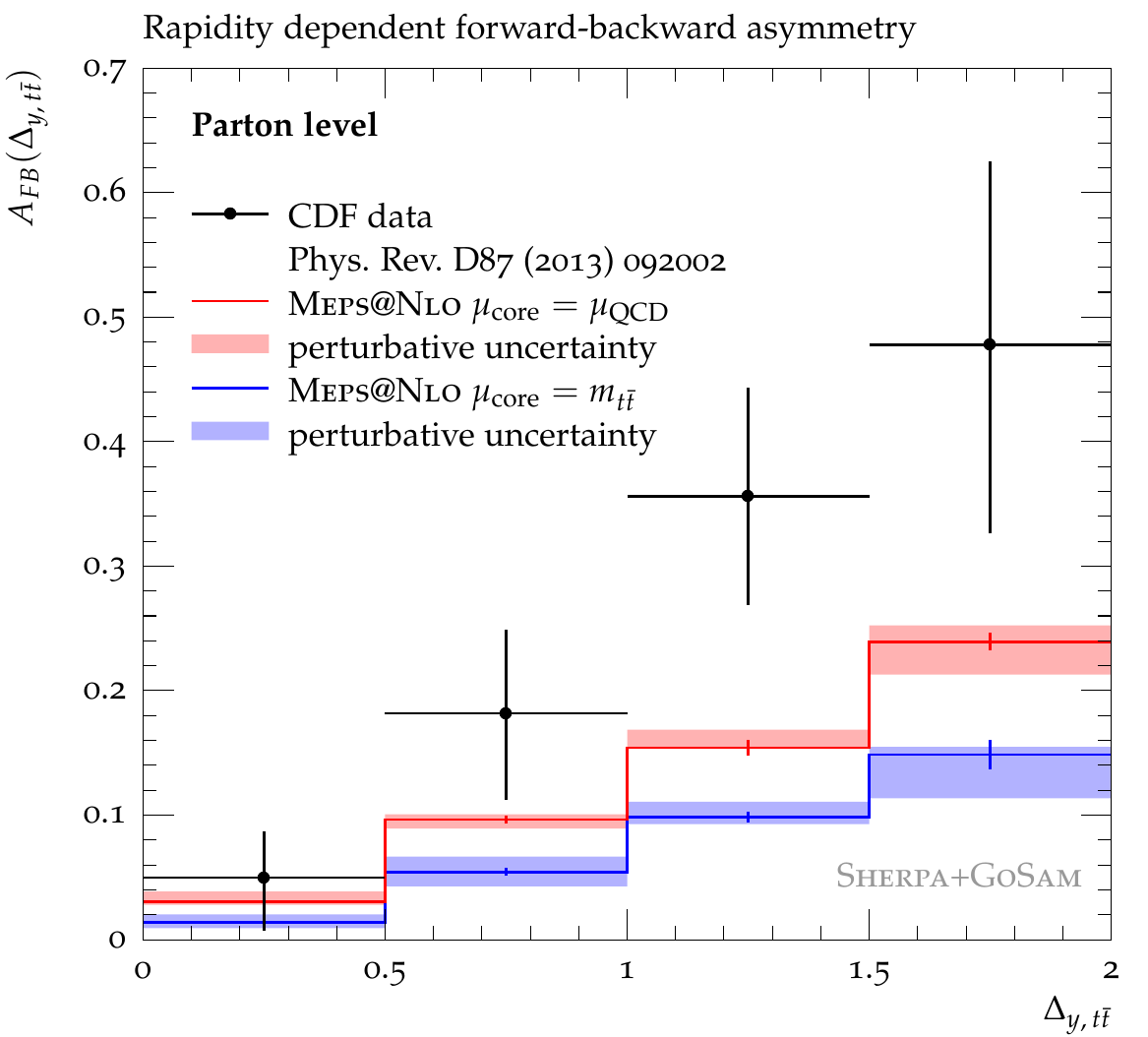}\hfill
    \includegraphics[width=0.47\textwidth]{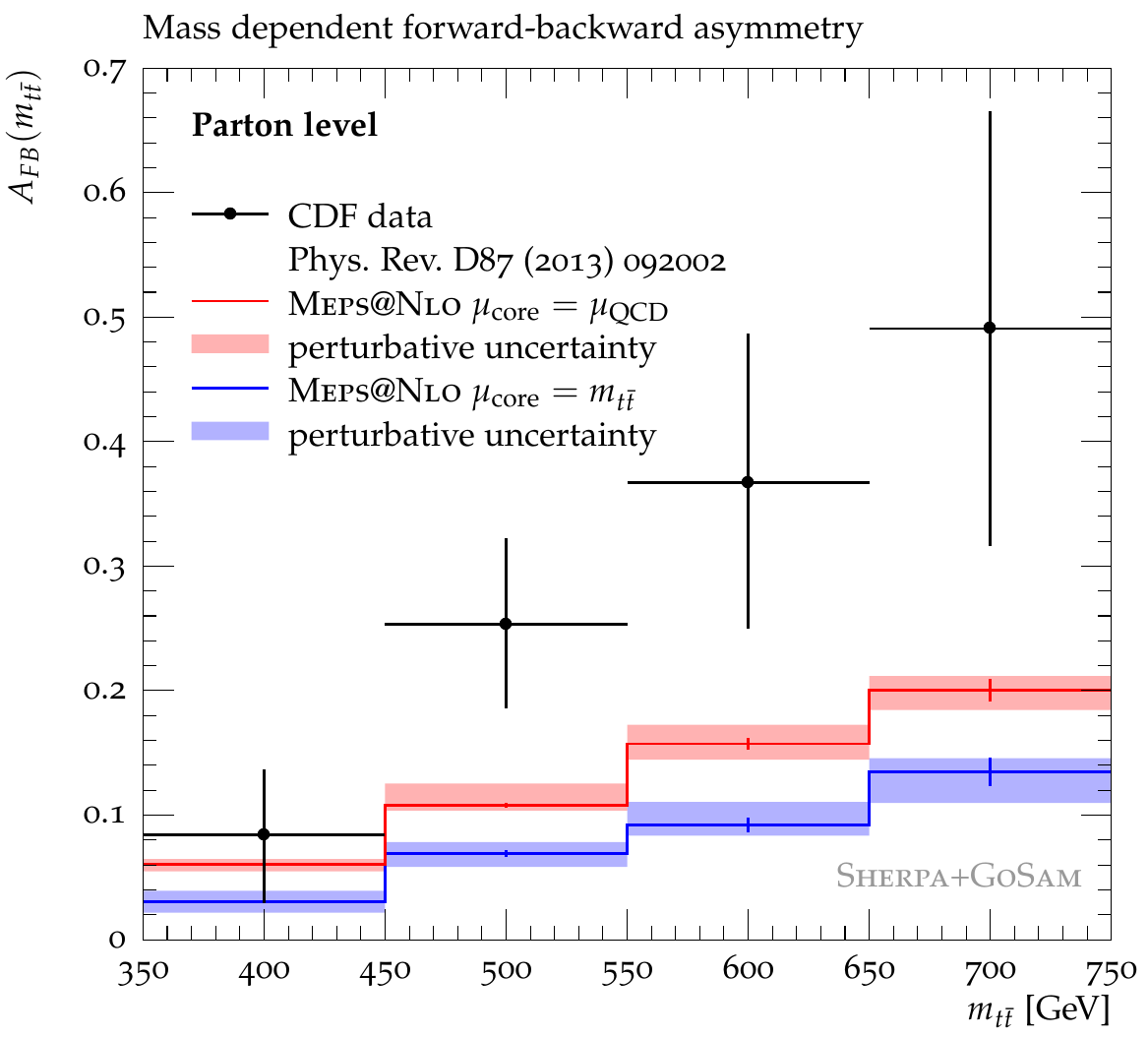}
  \end{center}
  \vspace*{-5mm}
  \caption
  {
    The $t\bar{t}$ asymmetry at the \Tevatron in dependence on the 
    transverse momentum (top), the rapidity separation (bottom left), 
    and the invariant mass (bottom right) of the $t\bar{t}$-pair 
    compared to \CDF data \cite{Aaltonen:2012it}.
    \label{fig:afb-asym}
  }
  \vspace*{-3mm}
\end{figure}

Fig.\ \ref{fig:afb-scales} shows the predictions of both scale choices 
for a standard observable, such as the top quark transverse momentum. They lead 
to consistent results with overlapping scale uncertainties. The 
perturbative convergence is slightly better for $\mu_\text{core}=
\mu_\text{QCD}$ as the \NLO uncertainty band is contained in the 
\LO uncertainty band. 

Fig.\ \ref{fig:afb-coherence} then highlights the importance of the 
inclusion of subleading colour terms in the resummation kernels of the 
\MCatNLO formulation of \cite{Hoeche:2011fd} as opposed to the 
$N_c\to\infty$ treatment utilised in conventional parton showers. While the 
effect on standard observables like the transverse momentum of the 
$t\bar{t}$-system is almost fully covered by the scale uncertainty of the 
resummation, its impact on its associated forward-backward asymmetry 
is profound.

Finally, Fig.\ \ref{fig:afb-asym} shows the forward-backward asymmetries 
wrt.\ to the transverse momentum, the rapidity difference and the invariant 
mass of the $t\bar{t}$-system for both core scale choices and the associated 
scale uncertainties. Data is well described for $A_\text{FB}(p_{\perp,t\bar{t}})$, 
while the description of $A_\text{FB}(\Delta_{y,t\bar{t}})$ and 
$A_\text{FB}(m_{t\bar{t}})$ is still poor. More important, however, 
is the fact that the difference between the two central scale choices is 
much larger than each individual uncertainty band. This roots in the fact 
that i) the forward-backward asymmetry is a ratio of observables, thus the 
effect of scale variations largely cancels, and ii) both scales behave 
differently wrt.\ forward or backward $t\bar{t}$ production configurations. 
Such effects have to be born in mind when evaluating the true theoretical 
uncertainties on these and similar observables.

\section{Conclusions}

The top quark forward--backward asymmetry at the \Tevatron has been 
analysed using a $p\bar{p}\to t\bar{t}+0,1\text{jets}$ next-to-leading order 
merged calculation. While this gives a good description of the asymmetry 
in dependence on the $t\bar{t}$-pair transverse momentum, the asymmetry 
in dependence on the $t\bar{t}$-pair rapidity difference and invariant 
mass, even taking into account additional EW effects, remains poor. 
Nonetheless, a consistent description of both the Sudakov region of the 
$p_{\perp,t\bar{t}}$ spectrum and the high $p_{\perp,t\bar{t}}$ region 
has been achieved. Furthermore, the uncertainty of the theory predictions 
has been reduced owing to the use of a next-to-leading order merged 
calculation as opposed to a leading order merged one. However, the variation 
resulting from using two distinct scale choices outsizes the variation 
due to shifting the individual scales by factors of 2. This has been 
demonstrated explicitly by employing two scales with different functional 
form. Last but not least, sub-leading colour terms in the first emission 
of the $t\bar{t}$ production process have been shown to have a large impact 
on the asymmetries.

\bibliographystyle{bib/amsunsrt_mod}
\bibliography{bib/journal}

\begin{thebibliography}{10}

\bibitem{Abazov:2007ab}
V.~Abazov et~al., D0 Collaboration collaboration, \emph{{First measurement of
  the forward-backward charge asymmetry in top quark pair production}},
  Phys.Rev.Lett. \textbf{100} (2008),
  \href{http://inspirehep.net/record/769720}{142002},
  [\href{http://arXiv.org/pdf/0712.0851}{{\tt arXiv:0712.0851}} [hep-ex]].
  \relax
 \relax
\bibitem{Aaltonen:2008hc}
T.~Aaltonen et~al., CDF Collaboration collaboration, \emph{{Forward-Backward
  Asymmetry in Top Quark Production in $p\bar{p}$ Collisions at $sqrt{s}=1.96$
  TeV}}, Phys.Rev.Lett. \textbf{101} (2008),
  \href{http://inspirehep.net/record/788356}{202001},
  [\href{http://arXiv.org/pdf/0806.2472}{{\tt arXiv:0806.2472}} [hep-ex]].
  \relax
 \relax
\bibitem{Aaltonen:2011kc}
T.~Aaltonen et~al., CDF Collaboration collaboration, \emph{{Evidence for a Mass
  Dependent Forward-Backward Asymmetry in Top Quark Pair Production}},
  Phys.Rev. \textbf{D83} (2011),
  \href{http://inspirehep.net/record/882996}{112003},
  [\href{http://arXiv.org/pdf/1101.0034}{{\tt arXiv:1101.0034}} [hep-ex]].
  \relax
 \relax
\bibitem{Abazov:2011rq}
V.~M. Abazov et~al., D0 Collaboration collaboration, \emph{{Forward-backward
  asymmetry in top quark-antiquark production}}, Phys.Rev. \textbf{D84} (2011),
  \href{http://inspirehep.net/record/920184}{112005},
  [\href{http://arXiv.org/pdf/1107.4995}{{\tt arXiv:1107.4995}} [hep-ex]].
  \relax
 \relax
\bibitem{Aaltonen:2012it}
T.~Aaltonen et~al., CDF collaboration, \emph{{Measurement of the top quark
  forward-backward production asymmetry and its dependence on event kinematic
  properties}}, Phys.Rev. \textbf{D87} (2013),
  \href{http://inspirehep.net/record/1198155}{092002},
  [\href{http://arXiv.org/pdf/1211.1003}{{\tt arXiv:1211.1003}} [hep-ex]].
  \relax
 \relax
\bibitem{Skands:2012mm}
P.~Skands, B.~Webber and J.~Winter, \emph{{QCD Coherence and the Top Quark
  Asymmetry}}, JHEP \textbf{1207} (2012),
  \href{http://inspirehep.net/record/1113762}{151},
  [\href{http://arXiv.org/pdf/1205.1466}{{\tt arXiv:1205.1466}} [hep-ph]].
  \relax
 \relax
\bibitem{Alioli:2011as}
S.~Alioli, S.-O. Moch and P.~Uwer, \emph{{Hadronic top-quark pair-production
  with one jet and parton showering}}, JHEP \textbf{1201} (2012),
  \href{http://www.slac.stanford.edu/spires/find/hep/www?eprint=1110.5251}{137},
   [\href{http://arXiv.org/pdf/1110.5251}{{\tt arXiv:1110.5251}} [hep-ph]].
  \relax
 \relax
\bibitem{Kardos:2011qa}
A.~Kardos, C.~Papadopoulos and Z.~Trocsanyi, \emph{{Top quark pair production
  in association with a jet with NLO parton showering}}, Phys.Lett.
  \textbf{B705} (2011), \href{http://inspirebeta.net/record/884386}{76--81},
  [\href{http://arXiv.org/pdf/1101.2672}{{\tt arXiv:1101.2672}} [hep-ph]].
  \relax
 \relax
\bibitem{Hoeche:2013mua}
S.~H{\"o}che, J.~Huang, G.~Luisoni, M.~Sch{\"o}nherr and J.~Winter, \emph{{Zero
  and one jet combined NLO analysis of the top quark forward-backward
  asymmetry}}, Phys.Rev. \textbf{D88} (2013),
  \href{http://inspirehep.net/record/1238288}{014040},
  [\href{http://arXiv.org/pdf/1306.2703}{{\tt arXiv:1306.2703}} [hep-ph]].
  \relax
 \relax
\bibitem{Hollik:2011ps}
W.~Hollik and D.~Pagani, \emph{{The electroweak contribution to the top quark
  forward-backward asymmetry at the Tevatron}}, Phys.Rev. \textbf{D84} (2011),
  \href{http://www.slac.stanford.edu/spires/find/hep/www?eprint=1107.2606}{093003},
   [\href{http://arXiv.org/pdf/1107.2606}{{\tt arXiv:1107.2606}} [hep-ph]].
  \relax
 \relax
\bibitem{Gleisberg:2008ta}
T.~Gleisberg, S.~H{\"o}che, F.~Krauss, M.~Sch\"{o}nherr, S.~Schumann,
  F.~Siegert and J.~Winter, \emph{{Event generation with \Sherpa 1.1}}, JHEP
  \textbf{02} (2009), \href{http://inspirebeta.net/record/803708}{007},
  [\href{http://arXiv.org/pdf/0811.4622}{{\tt arXiv:0811.4622}} [hep-ph]].
  \relax
 \relax
\bibitem{Krauss:2001iv}
F.~Krauss, R.~Kuhn and G.~Soff, \emph{{AMEGIC++ 1.0: A Matrix Element Generator
  In C++}}, JHEP \textbf{02} (2002),
  \href{http://www.slac.stanford.edu/spires/find/hep/www?eprint=hep-ph/0109036}{044},
   [\href{http://arXiv.org/pdf/hep-ph/0109036}{{\tt hep-ph/0109036}}]. \relax
 \relax
\bibitem{Gleisberg:2003ue}
T.~Gleisberg, F.~Krauss, K.~T. Matchev, A.~Sch{\"a}licke, S.~Schumann and
  G.~Soff, \emph{{Helicity formalism for spin-2 particles}}, JHEP \textbf{09}
  (2003),
  \href{http://www.slac.stanford.edu/spires/find/hep/www?eprint=hep-ph/0306182}{001},
   [\href{http://arXiv.org/pdf/hep-ph/0306182}{{\tt hep-ph/0306182}}]. \relax
 \relax
\bibitem{Gleisberg:2008fv}
T.~Gleisberg and S.~H{\"o}che, \emph{{Comix, a new matrix element generator}},
  JHEP \textbf{12} (2008), \href{http://inspirehep.net/record/793879}{039},
  [\href{http://arXiv.org/pdf/0808.3674}{{\tt arXiv:0808.3674}} [hep-ph]].
  \relax
 \relax
\bibitem{Catani:1996vz}
S.~Catani and M.~H. Seymour, \emph{{A general algorithm for calculating jet
  cross sections in NLO QCD}}, Nucl. Phys. \textbf{B485} (1997),
  \href{http://www.slac.stanford.edu/spires/find/hep/www?eprint=hep-ph/9605323}{291--419},
   [\href{http://arXiv.org/pdf/hep-ph/9605323}{{\tt hep-ph/9605323}}]. \relax
 \relax
\bibitem{Catani:2002hc}
S.~Catani, S.~Dittmaier, M.~H. Seymour and Z.~Trocsanyi, \emph{{The dipole
  formalism for next-to-leading order QCD calculations with massive partons}},
  Nucl. Phys. \textbf{B627} (2002),
  \href{http://www.slac.stanford.edu/spires/find/hep/www?eprint=hep-ph/0201036}{189--265},
   [\href{http://arXiv.org/pdf/hep-ph/0201036}{{\tt hep-ph/0201036}}]. \relax
 \relax
\bibitem{Gleisberg:2007md}
T.~Gleisberg and F.~Krauss, \emph{{Automating dipole subtraction for QCD NLO
  calculations}}, Eur. Phys. J. \textbf{C53} (2008),
  \href{http://www.slac.stanford.edu/spires/find/hep/www?eprint=arXiv:0709.2881}{501--523},
   [\href{http://arXiv.org/pdf/0709.2881}{{\tt arXiv:0709.2881}} [hep-ph]].
  \relax
 \relax
\bibitem{Schumann:2007mg}
S.~Schumann and F.~Krauss, \emph{{A parton shower algorithm based on
  Catani-Seymour dipole factorisation}}, JHEP \textbf{03} (2008),
  \href{http://www.slac.stanford.edu/spires/find/hep/www?eprint=arXiv:0709.1027}{038},
   [\href{http://arXiv.org/pdf/0709.1027}{{\tt arXiv:0709.1027}} [hep-ph]].
  \relax
 \relax
\bibitem{Cullen:2011ac}
G.~Cullen, N.~Greiner, G.~Heinrich, G.~Luisoni, P.~Mastrolia, G.~Ossola,
  T.~Reiter and F.~Tramontano, \emph{{Automated One-Loop Calculations with
  GoSam}}, Eur.Phys.J. \textbf{C72} (2012),
  \href{http://www.slac.stanford.edu/spires/find/hep/www?rawcmd=FIND+EPRINT+1111.2034}{1889},
   [\href{http://arXiv.org/pdf/1111.2034}{{\tt arXiv:1111.2034}} [hep-ph]].
  \relax
 \relax
\bibitem{Mastrolia:2010nb}
P.~Mastrolia, G.~Ossola, T.~Reiter and F.~Tramontano, \emph{{Scattering
  AMplitudes from Unitarity-based Reduction Algorithm at the Integrand-level}},
  JHEP \textbf{1008} (2010), \href{http://inspirebeta.net/record/856935}{080},
  [\href{http://arXiv.org/pdf/1006.0710}{{\tt arXiv:1006.0710}} [hep-ph]].
  \relax
 \relax
\bibitem{Binoth:2010xt}
T.~Binoth et~al., \emph{{A proposal for a standard interface between Monte
  Carlo tools and one-loop programs}}, Comput. Phys. Commun. \textbf{181}
  (2010), \href{http://inspirehep.net/record/842428}{1612--1622},
  [\href{http://arXiv.org/pdf/1001.1307}{{\tt arXiv:1001.1307}} [hep-ph]].
  \relax
 \relax
\bibitem{Alioli:2013nda}
\href{http://www.slac.stanford.edu/spires/find/hep/www?eprint=1308.3462}{S.~Alioli
  et~al.}, \emph{{Update of the Binoth Les Houches Accord for a standard
  interface between Monte Carlo tools and one-loop programs}},
  \href{http://arXiv.org/pdf/1308.3462}{{\tt arXiv:1308.3462}} [hep-ph]. \relax
 \relax
\bibitem{Martin:2009iq}
A.~D. Martin, W.~J. Stirling, R.~S. Thorne and G.~Watt, \emph{{Parton
  distributions for the LHC}}, Eur. Phys. J. \textbf{C63} (2009),
  \href{http://www-spires.dur.ac.uk/spires/find/hep/www?eprint=arXiv:0901.0002}{189--295},
   [\href{http://arXiv.org/pdf/0901.0002}{{\tt arXiv:0901.0002}} [hep-ph]].
  \relax
 \relax
\bibitem{Hoeche:2009rj}
S.~H{\"o}che, F.~Krauss, S.~Schumann and F.~Siegert, \emph{{QCD matrix elements
  and truncated showers}}, JHEP \textbf{05} (2009),
  \href{http://www.slac.stanford.edu/spires/find/hep/www?eprint=arXiv:0903.1219}{053},
   [\href{http://arXiv.org/pdf/0903.1219}{{\tt arXiv:0903.1219}} [hep-ph]].
  \relax
 \relax
\bibitem{Hoeche:2010kg}
S.~H{\"o}che, F.~Krauss, M.~Sch{\"o}nherr and F.~Siegert, \emph{{NLO matrix
  elements and truncated showers}}, JHEP \textbf{08} (2011),
  \href{http://www.slac.stanford.edu/spires/find/hep/www?eprint=arXiv:1009.1127}{123},
   [\href{http://arXiv.org/pdf/1009.1127}{{\tt arXiv:1009.1127}} [hep-ph]].
  \relax
 \relax
\bibitem{Hoeche:2012yf}
S.~H{\"o}che, F.~Krauss, M.~Sch{\"o}nherr and F.~Siegert, \emph{{QCD matrix
  elements + parton showers: The NLO case}}, JHEP \textbf{1304} (2013),
  \href{http://inspirehep.net/record/1123387}{027},
  [\href{http://arXiv.org/pdf/1207.5030}{{\tt arXiv:1207.5030}} [hep-ph]].
  \relax
 \relax
\bibitem{Gehrmann:2012yg}
T.~Gehrmann, S.~H{\"o}che, F.~Krauss, M.~Sch{\"o}nherr and F.~Siegert,
  \emph{{NLO QCD matrix elements + parton showers in $e^+e^-\to$hadrons}}, JHEP
  \textbf{1301} (2013),
  \href{http://www.slac.stanford.edu/spires/find/hep/www?eprint=1207.5031}{144},
   [\href{http://arXiv.org/pdf/1207.5031}{{\tt arXiv:1207.5031}} [hep-ph]].
  \relax
 \relax
\bibitem{Hoeche:2011fd}
S.~H{\"o}che, F.~Krauss, M.~Sch{\"o}nherr and F.~Siegert, \emph{{A critical
  appraisal of NLO+PS matching methods}}, JHEP \textbf{09} (2012),
  \href{http://inspirehep.net/record/944643}{049},
  [\href{http://arXiv.org/pdf/1111.1220}{{\tt arXiv:1111.1220}} [hep-ph]].
  \relax
 \relax
\bibitem{Hoeche:2012ft}
S.~H{\"o}che, F.~Krauss, M.~Sch{\"o}nherr and F.~Siegert, \emph{{W+n-jet
  predictions with MC@NLO in Sherpa}}, Phys.Rev.Lett. \textbf{110} (2013),
  \href{http://inspirehep.net/record/1086175}{052001},
  [\href{http://arXiv.org/pdf/1201.5882}{{\tt arXiv:1201.5882}} [hep-ph]].
  \relax
 \relax
\bibitem{Hoeche:2012fm}
S.~H{\"o}che and M.~Sch{\"o}nherr, \emph{{Uncertainties in next-to-leading
  order plus parton shower matched simulations of inclusive jet and dijet
  production}}, Phys.Rev. \textbf{D86} (2012),
  \href{http://inspirehep.net/record/1127523}{094042},
  [\href{http://arXiv.org/pdf/1208.2815}{{\tt arXiv:1208.2815}} [hep-ph]].
  \relax
 \relax
\bibitem{Cascioli:2013era}
\href{http://www.slac.stanford.edu/spires/find/hep/www?eprint=1309.5912}{F.~Cascioli,
  P.~Maierhoefer, N.~Moretti, S.~Pozzorini and F.~Siegert}, \emph{{NLO matching
  for $t\bar{t}b\bar{b}$ production with massive $b$-quarks}},
  \href{http://arXiv.org/pdf/1309.5912}{{\tt arXiv:1309.5912}} [hep-ph]. \relax
 \relax
\end{thebibliography}

\end{document}